\documentclass[aps,prl,twocolumn,superscriptaddress,showpacs]{revtex4}

\usepackage{graphicx}

\begin{document}

\title{Magnetic Field Effect on the Superconducting Magnetic Gap of
Nd$_{1.85}$Ce$_{0.15}$CuO$_4$}

\author{E.M.~Motoyama} \affiliation{Department of Physics, Stanford
University, Stanford, California 94305}

\author{P.K.~Mang} \affiliation{Department of Applied Physics, Stanford
University, Stanford, California 94305}

\author{D.~Petitgrand} \affiliation{Laboratoire L\'{e}on Brillouin
(CEA-CNRS), CEA-Saclay, 91191 Gif-sur-Yvette Cedex, France}

\author{G.~Yu} \affiliation{Department of Physics, Stanford University,
Stanford, California 94305}

\author{O.P.~Vajk} \affiliation{NIST Center for Neutron Research,
National Institute of Standards and Technology, Gaithersburg, Maryland
20899}

\author{I.M.~Vishik} \affiliation{Department of Physics, Stanford
University, Stanford, California 94305}

\author{M.~Greven} \email[]{greven@stanford.edu} \affiliation{Department
of Applied Physics, Stanford University, Stanford, California 94305}
\affiliation{Stanford Synchrotron Radiation Laboratory, Stanford,
California 94309}

\date{\today}

\begin{abstract} Inelastic neutron scattering measurements on the
archetypical electron-doped material Nd$_{1.85}$Ce$_{0.15}$CuO$_4$ up to
high relative magnetic field strength, $H / H_{c2} \sim 50\%$, reveal a
simple linear magnetic-field effect on the superconducting magnetic gap
and the absence of field-induced in-gap states. The extrapolated
gap-closing field value is consistent with the upper critical field
$H_{c2}$, and the high-field response resembles that of the paramagnetic
normal state. \end{abstract}


\maketitle

A pivotal problem in condensed matter physics during the past two
decades has been the determination of the ground state phase diagram of
the high-temperature superconductors as a function of carrier density
and magnetic field. Knowledge about the ground state that competes with
the superconducting (SC) phase may be crucial to understanding the
mechanism of superconductivity; this can be studied by suppressing
superconductivity in a strong magnetic field, which leads to the
formation of non-SC vortex regions embedded in the otherwise SC material
\cite{Huebener01}.

The high-$T_c$ superconductors are divided into two classes, depending
on whether they are formed by doping electrons or holes into the
antiferromagnetic (AF) Mott insulator parent compounds
\cite{KastnerRMP98}.  In the hole-doped materials, the low-energy
magnetic response is incommensurate and manifests itself in neutron
scattering as four peaks situated symmetrically around the AF zone
center $(\pi,\pi)$ \cite{CheongYamadaMook}. Recent neutron-scattering
studies in a magnetic field indicate that the non-SC ground state may be
magnetically ordered
\cite{LakeScience01,LakeNature02,KhaykovichPRB02,TranquadaPRB04}.
Specifically, in underdoped La$_{2-x}$Sr$_x$CuO$_4$ (LSCO) and
La$_2$CuO$_{4+\delta}$, applying a magnetic field enhances the
incommensurate spin density wave (SDW) order already present in the
system \cite{LakeNature02,KhaykovichPRB02}. In optimally and
slightly-overdoped LSCO, for which the magnetic signal is gapped (i.e.,
no static order) \cite{YamadaPRL95}, a magnetic field induces new
excitations below the gap energy
\cite{LakeScience01,TranquadaPRB04,GilardiEPL04}. The phase diagram thus
contains a SC phase and a phase with coexisting SC and SDW order
\cite{DemlerZhang}. The overdoped system lies in the SC phase, and a
magnetic field pushes it towards the SC+SDW phase. The underdoped system
already lies in the SC+SDW phase at zero field. There exists a
zero-field quantum critical point between these two phases.
Interestingly, recent results at intermediate doping show that a
relatively low magnetic field of $H = 3\,\mathrm{T}$ pushes the system
from the SC into the SC+SDW phase \cite{KhaykovichPRB05}.

Upon doping, the electron-doped compounds exhibit a particularly robust
neighboring AF phase in zero magnetic field \cite{MangPRL04}, and the
magnetic response remains commensurate at $(\pi,\pi)$
\cite{YamadaPRL03}. One might expect the phase diagram to be similar to
the hole-doped case, with superconductivity coexisting with commensurate
AF order instead of incommensurate SDW order. Initial experiments on
Nd$_{1.85}$Ce$_{0.15}$CuO$_4$ agreed with this na\"{\i}ve picture: a
magnetic field perpendicular to the CuO$_2$ sheets was found to induce
an elastic signal at $(\pi,\pi)$ \cite{KangNature03}. However, it was
subsequently shown that this signal is spurious: it is due to the
paramagnetic response of the epitaxial secondary phase $(\mathrm{Nd},
\mathrm{Ce})_2\mathrm{O}_3$ \cite{MangNature03,MangPRB04}. Consequently,
the important question of the nature of the field-induced ground state
on the electron-doped side of the phase diagram has remained open. We
note that large crystals of SC NCCO typically contain a second spurious
signal due to regions that have not been fully oxygen-reduced. These AF
NCCO remnants have a three-dimensional or quasi-two-dimensional spatial
extent, and they primarily contribute to the elastic response
\cite{MangPRB04,UefujiPhysicaC02}.  While these chemical properties
prevent a search for any genuine elastic signal, they do not prevent
measurements of the inelastic response from the SC majority volume
fraction. Indeed, a careful inelastic neutron scattering measurement on
NCCO has clearly revealed the SC magnetic gap below $T_c$
\cite{YamadaPRL03,PetitgrandPhysica04}.

In this Letter, we use inelastic neutron scattering to determine the
magnetic field effect on this SC gap. For the hole-doped materials, the
upper critical field at which superconductivity is completely destroyed
is $50\,\mathrm{T}$ or greater \cite{WangScience03}, and hence
prohibitively large for neutron scattering experiments. For the
electron-doped materials, on the other hand, $H_{c2}$ is relatively low
($\sim 10$ T) \cite{WangScience03}, which has allowed us to observe a
field-effect on the SC magnetic gap in NCCO up to high values of
relative magnetic field strength. We find that applying a magnetic field
causes a rigid shift of the gap profile to lower energies. This
contrasts the case of optimally-doped and over-doped LSCO, in which a
magnetic field introduces in-gap states. A complementary zero-field
measurement on non-SC
Nd$_{1.85}$Ce$_{0.15}$Cu$_{0.985}$Ni$_{0.015}$O$_4$ demonstrates that
the low-temperature spin correlations remain finite. Our findings imply
a uniform low-energy magnetic response for fields below $H_{c2}$, and
they are consistent with a paramagnetic ground state in the absence of
superconductivity.

\begin{figure}[b] \includegraphics[width=2.6in]{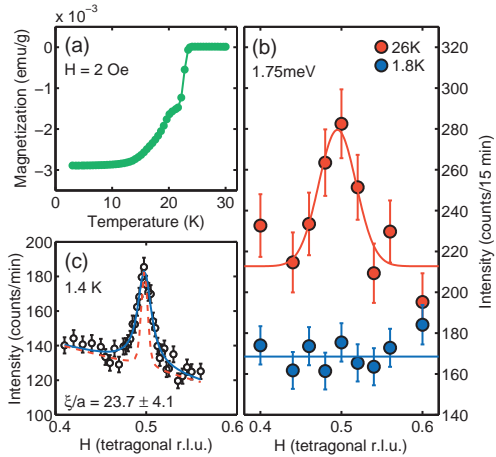}
\caption{\label{fig1} (color online). (a) SQUID magnetometry of the
whole Nd$_{1.85}$Ce$_{0.15}$CuO$_4$ crystal used in the neutron
scattering experiment showing an onset $T_c$ of $22\,\mathrm{K}$. (b)
$(H,1-H,0)$ scans through $(1/2,1/2,0)$ at an energy transfer of $\omega
= 1.75\,\mathrm{meV}$ in Nd$_{1.85}$Ce$_{0.15}$CuO$_4$ above ($T =
26\,\mathrm{K}$) and well below ($T = 1.8\,\mathrm{K}$) $T_c$. (c)
Determination of the instantaneous spin correlation length $\xi$ at $T =
1.4\,\mathrm{K}$ in non-SC
Nd$_{1.85}$Ce$_{0.15}$Cu$_{0.985}$Ni$_{0.015}$O$_4$ using the
energy-integrating two-axis scan \cite{MangPRL04}. The data are fit
(continuous line) to a Lorentzian convoluted with the calculated
resolution function (dashed line).} \end{figure}

The inelastic neutron scattering experiment was carried out on the 4F2
triple-axis spectrometer at the Laboratoire L\'{e}on Brillouin in
Saclay, France, with final neutron energy of 14.7 meV and collimations
of $60'$-open-sample-open-open, and with a horizontally focusing
monochromator and a vertically focusing analyzer. The
Nd$_{1.85}$Ce$_{0.15}$CuO$_4$ crystal (mass $5.0\,\mathrm{g}$) was grown
at Stanford University in an oxygen-flow at a pressure of
$4\,\mathrm{atm}$ using the traveling-solvent floating-zone technique,
and subsequently annealed for 10 hours at $970{^\circ}\mathrm{C}$ in
argon, followed by $500{^\circ}\mathrm{C}$ for 20 hours in oxygen
\cite{MangPRB04}. As shown in Fig.~\ref{fig1}(a), the crystal exhibits
an onset $T_c$ of $22\,\mathrm{K}$. A second, non-SC crystal of
Nd$_{1.85}$Ce$_{0.15}$Cu$_{0.985}$Ni$_{0.015}$O$_4$ (mass
$3.1\,\mathrm{g}$) was prepared by the same method and studied in zero
field in two-axis mode \cite{MangPRL04}. These two-axis data were taken
on the BT9 spectrometer at the NIST Center for Neutron Research, with an
incident neutron energy of 14.7 meV and collimations
$40'$-$46.7'$-sample-$10'$. Inductively coupled plasma (ICP) analysis
indicates that the composition of the Ni-doped crystal is close to its
specified nominal values, while the Ce concentration of the Ni-free
crystal is less uniform and closer to $x=0.16$.

NCCO has a tetragonal unit cell (space group $I4/mmm$) with
low-temperature lattice constants $a = 3.94\,\textrm{\AA}$ and $c =
12.09\,\textrm{\AA}$ for $x=0.15$. We represent wavevectors as $(H,K,L)$
in reciprocal lattice units (r.l.u.), where $\mathbf{Q} = (2\pi H/a,
2\pi K/a, 2\pi L/c)$. The AF zone center at $(1/2,1/2,0)$ is also
represented as $(\pi,\pi)$ (i.e., $a\equiv 1$). The inelastic
experiments on SC NCCO were performed in the $(H,K,0)$ geometry, in
which the crystal $c$-axis is perpendicular to the scattering plane. The
magnetic field was applied vertically in the $c$ direction.  We first
establish, in Fig.~\ref{fig1}(b), that the magnetic excitations at
$(\pi,\pi)$ are indeed gapped below $T_c$. Shown are transverse
$Q$-space scans centered at $(H,K,L) = (1/2,1/2,0)$ at an energy
transfer of $\omega = 1.75\,\mathrm{meV}$. As expected, there is a clear
peak above $T_c$ at $T = 26\,\mathrm{K}$, while the signal is suppressed
below $T_c$ at $T = 1.8\,\mathrm{K}$.

\begin{figure}[b] \includegraphics[width=2.6in]{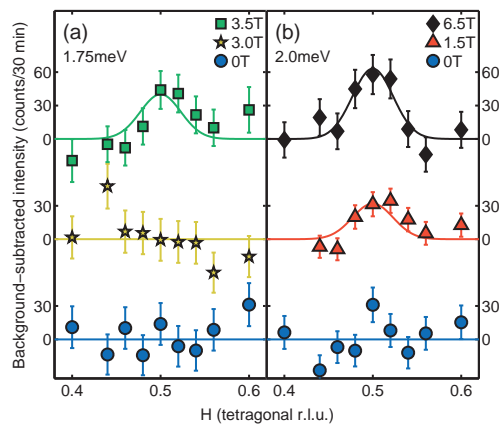}
\caption{\label{fig2} (color online). (a) Transverse scans $(H,1-H,0)$
about the AF zone center $(1/2,1/2,0)$ at an energy transfer of $\omega
= 1.75\,\mathrm{meV}$ and (b) $\omega = 2.0\,\mathrm{meV}$. Before each
scan, the sample was field-cooled from above $T_c$ to $T =
1.8\,\mathrm{K}$. Typical counting time is 30 minutes per point.}
\end{figure}

In Fig.~\ref{fig2}, we see how an applied magnetic field can cause this
suppressed signal to reappear. For every scan at a new field, the sample
was first heated above $T_c$ and then cooled in the new field back down
to $T = 1.8\,\mathrm{K}$; this was done to ensure a macroscopically
uniform internal field. At an energy transfer of $\omega =
1.75\,\mathrm{meV}$ [Fig.~\ref{fig2}(a)], the magnetic excitations are
completely suppressed up to $H = 3\,\mathrm{T}$, and reemerge at $H =
3.5\,\mathrm{T}$. A similar behavior is seen at the slightly higher
energy transfer of $\omega = 2.0\,\mathrm{meV}$ [Fig.~\ref{fig2}(b)]. In
this case, the peak is seen to reappear at a lower field of $H =
1.5\,\mathrm{T}$.

\begin{figure}[t] \includegraphics[width=2.1in]{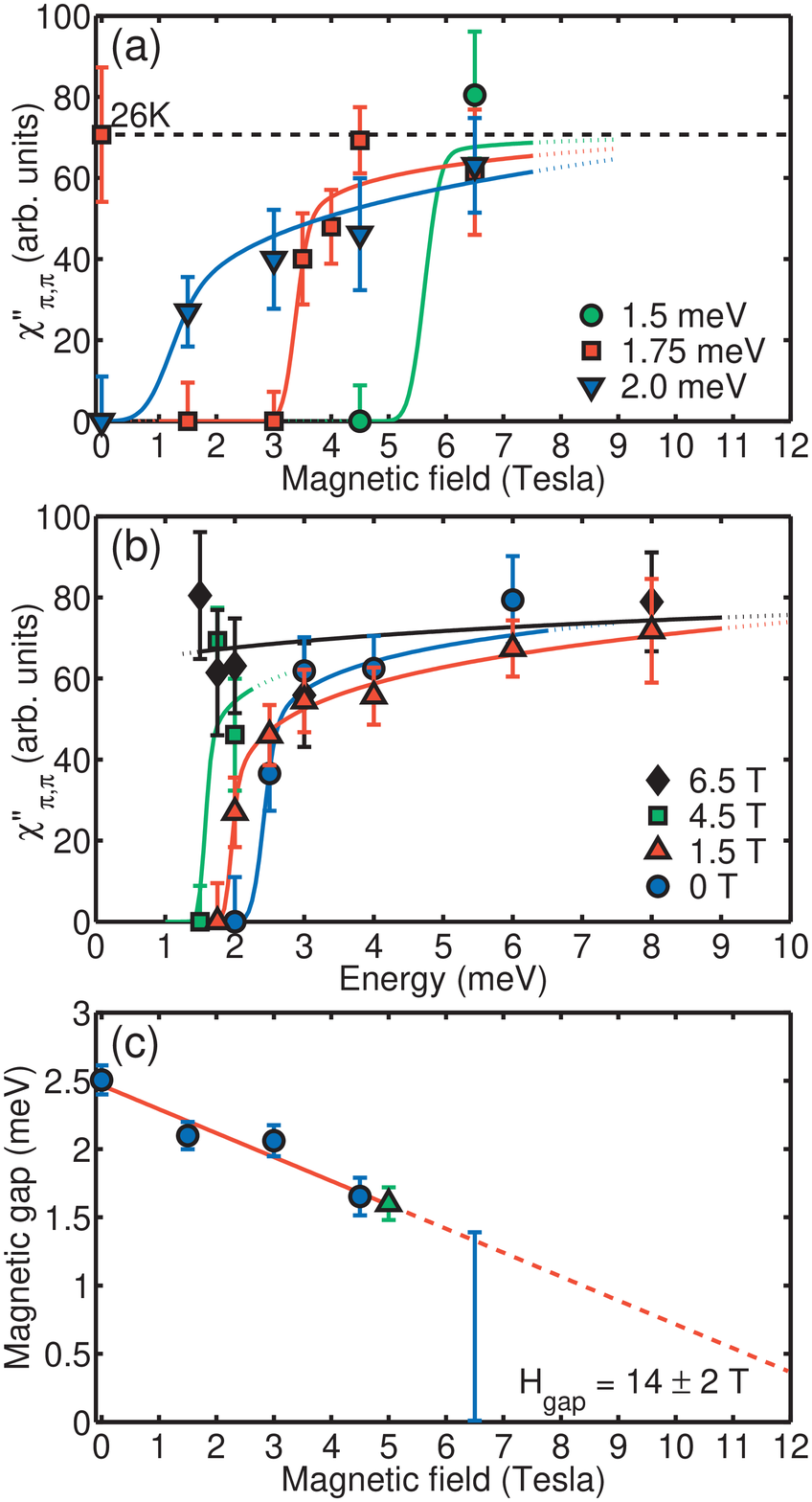}
\caption{\label{fig3} (color online). (a) Dynamic susceptibility
$\chi''$ at ($\pi,\pi$) as a function of field at several energies. All
data are at $T = 1.8\,\mathrm{K}$ except for the zero-field datum at $T
= 26\,\mathrm{K}$. Curves are guides to the eye. (b) The
field-dependence of the magnetic excitation spectrum $\chi''_{\pi,\pi}$
versus energy at $T = 1.8\,\mathrm{K}$. Curves are guides to the eye.
(c) Evolution of the magnetic gap (half-maximum energy) as a function of
field. The relation is linear, and extrapolates to a gap-closing
critical field of $H_\mathrm{gap} = 14 \pm 2\,\mathrm{T}$. The triangle
represents data taken at $H = 5\,\mathrm{T}$ upon zero-field cooling.
Although field-cooling is preferred (resulting in a uniform internal
field), our result is independent of the cooling method, as might be
expected at such high magnetic fields. The vertical bar at $H =
6.5\,\mathrm{T}$ reflects our limited knowledge that the gap energy is
less than $1.5\,\mathrm{meV}$.  } \end{figure}

We summarize these field-dependence results in Fig.~\ref{fig3}(a). Here
we plot the signal strength (written as the imaginary part of the
dynamic susceptibility, $\chi''$) as a function of magnetic field for
the two energy transfers, as well as for $\omega = 1.5\,\mathrm{meV}$.
The signal strength (corrected for the Bose factor) above $T_c$ at $T =
26\,\mathrm{K}$ for $\omega = 1.75\,\mathrm{meV}$ is shown as a
horizontal dashed line. We see that as the field is increased, the
signal strength approaches this normal-state value.

We plot the magnetic excitation spectrum at several field values in
Fig.~\ref{fig3}(b). The zero-field gap energy in our sample is
$2.5\,\mathrm{meV}$, slightly smaller than in previous work
\cite{PetitgrandPhysica04}, in accordance with our somewhat lower $T_c$.
As the magnetic field is increased, the gap profile shifts rigidly to
lower energies. In particular, note that the signal at lower energies
remains zero while the signal is restored by the magnetic field at
slightly higher energies. At our maximum field of $H = 6.5\,\mathrm{T}$,
we can no longer discern the gap. Due to a strong increase of the
background level at lower energies and the non-zero energy resolution of
$1.3\,\mathrm{meV}$ (FWHM), we were unable to make measurements below
$\omega = 1.5\,\mathrm{meV}$. The strong field-induced background
results from the spurious $(\mathrm{Nd}, \mathrm{Ce})_2\mathrm{O}_3$
magnetic elastic signal and from intrinsic paramagnetism of the
spin-polarized Nd subsystem of NCCO.

In Fig.~\ref{fig3}(c), we plot the gap energy as a function of field.
The magnetic gap decreases linearly with field, and it completely
collapses at an extrapolated value of $H_\mathrm{gap} = 14(2)
\,\mathrm{T}$, consistent with upper critical field of $H_{c2} \approx
\textrm{10--12}\,\mathrm{T}$ \cite{WangScience03}. The gapped spectrum
of NCCO undergoes a rigid shift towards lower energies as the magnetic
field is increased, which is in stark contrast to the formation of
in-gap states in optimally and slightly overdoped LSCO
\cite{LakeScience01,TranquadaPRB04}. Because measurements below
$1.5\,\mathrm{meV}$ have not been possible, it is natural to ask whether
the formation of some in-gap states in NCCO could be hidden below this
energy. However, we emphasize that the signal strength at
$1.5\,\mathrm{meV}$ remains zero up until $4.5\,\mathrm{T}$ [Fig. 3(a)].
 Since our energy resolution is $1.3\,\mathrm{meV}$ (FWHM), our
experiment is sensitive to any in-gap intensity down to very low
energies.

Quite generally, a magnetic field suppresses superconductivity by
orbital pair-breaking of Cooper pairs in the SC state and by lowering
the relative energy of the normal state via Pauli paramagnetism of the
electron spins \cite{Bianchi02}. In the orbital pair-breaking limit, the
expected zero-temperature critical field in NCCO is $H_\mathrm{orb}
\approx 9\,\mathrm{T}$, estimated from the slope of $H_{c2}(T)$ at $T_c$
\cite{WangScience03,Bianchi02}. In the Pauli limit, the critical field
is given by $H_p = \Delta_0/(\sqrt2 \mu_B)$. From measurements of the
the electronic gap $\Delta_0$ \cite{Matsui05,Qazilbash05}, we find $H_p
\sim \textrm{30--45}\,\mathrm{T}$. Since $H_\mathrm{gap} \sim H_{c2}
\sim H_\mathrm{orb}$, the response of our sample to a magnetic field is
dominated by orbital effects, namely the balance between the electron
kinetic energy and the condensation energy of the Cooper pairs.

We note that our results are not consistent with a singlet-triplet gap,
since the Zeeman effect would require a field of more than
$20\,\mathrm{T}$ to close the gap.  On the other hand, the observed
simple linear field dependence of the magnetic gap is what one obtains
in a na\"{\i}ve BCS picture of a spatially uniform response. In such a
picture, the magnetic gap is proportional to the SC electronic gap,
which in turn is proportional to $T_c$. According to vortex Nernst
effect measurements, $T_c$ decreases linearly as a function of field (at
least at lower fields) \cite{WangScience03}. This implies a linear
decrease in the SC electronic gap, as recently measured by Raman
spectroscopy \cite{Qazilbash05}.  What we have discovered is that the SC
magnetic gap also decreases linearly as a function of field. It is worth
noting that our measured magnetic gap of $2.5\,\mathrm{meV}$ is much
smaller than twice the electronic gap \cite{Matsui05,Qazilbash05}.  This
suggests a non-trivial relationship between the electronic and magnetic
gaps.  A possible theoretical explanation for this observation has been
put forward recently \cite{OnufrievaPRL04}.

While the picture of a uniform response is consistent with our data, we
do know that the {\em electronic} response is not uniform. The size of
the non-SC vortex cores (the SC coherence length) has been estimated to
be $58\,\textrm{\AA}$ in NCCO \cite{WangScience03}. However, we find
that the field-induced low-temperature response remains
momentum-resolution limited, which implies that the dynamic magnetic
correlations are long-range (at least $200\,\textrm{\AA}$), spanning
both vortex-core and SC regions. This provides further evidence that the
{\em magnetic} low-energy response of NCCO is uniform.

The in-gap signal that is present in LSCO, but absent in NCCO, has been
attributed in $SO(5)$ theory to ``bound states'' in the vortex cores. In
this theory, the non-SC vortex cores act as attractive potentials for
magnetic excitations \cite{ArovasPRL97,HuJPCS02}. The dynamics of the AF
fluctuations can be described by a Schr\"{o}dinger-like equation, and
the in-gap signal seen in overdoped LSCO corresponds to the presence of
magnetic bound states in the vortex-core potentials. The absence of an
in-gap signal and the linear field dependence of the gap found in our
study of NCCO is consistent with the absence of such bound states
\cite{HuJPCS02}. Alternatively, the in-gap intensity in LSCO has been
attributed to the proximity of the SC to SC+SDW quantum phase transition
\cite{DemlerZhang}.  In this context, NCCO corresponds to a new region
of the phase diagram, far from such a phase transition.  Our results
indicate that unlike in the hole-doped case, there is no transition to
magnetic order before superconductivity disappears.

Our results point to a picture in which the non-SC ground state at
fields above $H_{c2}$ does not possess magnetic order, but is a
paramagnet with AF fluctuations. The first piece of evidence is that, in
NCCO, applying a magnetic field and increasing temperature have similar
effects, and the gap does not appear to close until SC is completely
suppressed \cite{YamadaPRL03}. Moreover, the signal strength seen at
high magnetic fields equals that in the normal state just above $T_c$.
All of this indicates that the non-SC ground state beyond $H_{c2}$
resembles the paramagnetic ÒnormalÓ state above $T_c$.

A second piece of evidence comes from our complementary study of
Ni-doped NCCO. Ni-doping is an alternative method of suppressing
superconductivity. Upon substituting only about 1\% of Cu with Ni,
superconductivity in NCCO is completely suppressed
\cite{TarasconSugiyamaJayaram}. We have performed zero-field
measurements of the instantaneous spin correlation length in an
oxygen-reduced non-SC sample of 1.5\% Ni-doped NCCO.
Figure~\ref{fig1}(c) shows that this system at low temperatures has a
finite correlation length: the momentum-space peaks are considerably
broader than the experimental resolution. Clearly, the non-SC ground
state in NCCO, induced by Ni-doping, does not have long-range magnetic
order.

The absence of magnetic-field-induced in-gap states in
Nd$_{1.85}$Ce$_{0.15}$CuO$_4$ and the likely absence of field-induced
magnetic order signify an important difference between the
electron-doped and hole-doped cuprates; the competing spin (and charge)
density wave order (often referred to as ``stripes'') observed in
hole-doped superconductors, especially in materials derived from the
high-$T_c$ parent compound La$_2$CuO$_4$, prohibits an unobstructed
study of the antiferromagnetically correlated superconductor due to the
presence of a nearby quantum critical point. This complication appears
to be avoided by the electron-doped materials, which possess the
additional experimental advantage of a relatively low upper critical
field.

\begin{acknowledgments} We would like to acknowledge helpful discussions
with G. Aeppli, W.J.L. Buyers, R.S. Markiewicz, F. Onufrieva, P. Pfeuty,
J. Zaanen, and S.C. Zhang. The work at Stanford University was supported
by DOE under Contracts No. DE-FG03-99ER45773 and No. DE-AC03-76SF00515,
and by NSF DMR 9985067. E.M.M. acknowledges support through the NSF
Graduate program. \end{acknowledgments}

\bibliography{emdm}

\end{document}